\documentclass[twocolumn]{aastex63}
\usepackage{verbatim}
\usepackage{comment} 
\usepackage{calrsfs}
\usepackage{amsmath}
\DeclareMathAlphabet{\pazocal}{OMS}{zplm}{m}{n}

\makeatletter
\def\@hex@@Hex#1%
 {\if a#1A\else \if b#1B\else \if c#1C\else \if d#1D\else
  \if e#1E\else \if f#1F\else #1\fi\fi\fi\fi\fi\fi \@hex@Hex}
\makeatother
\definecolor{afcolor}{HTML}{b3443c}

\received{August 25, 2025}
\accepted{\today}

\shorttitle{Star formation efficiency}
\shortauthors{Ferrara, Manzoni \& Ntormousi}
\graphicspath{{./}{figures/}}

\begin{document}



\def\be{\begin{equation}}
\def\ee{\end{equation}}
\newcommand\code[1]{\textsc{\MakeLowercase{#1}}}
\newcommand\quotesingle[1]{`{#1}'}
\newcommand\quotes[1]{``{#1}"}
\def\gsim{\lower.5ex\hbox{\gtsima}} 
\def\lsim{\lower.5ex\hbox{\ltsima}} 
\def\gtsima{$\; \buildrel > \over \sim \;$} 
\def\ltsima{$\; \buildrel < \over \sim \;$} \def\gsim{\lower.5ex\hbox{\gtsima}} 
\def\lsim{\lower.5ex\hbox{\ltsima}} 
\def\simgt{\lower.5ex\hbox{\gtsima}} 
\def\simlt{\lower.5ex\hbox{\ltsima}}

\def\msun{{\rm M}_{\odot}}
\def\lsun{{\rm L}_{\odot}}
\def\dsun{{\cal D}_{\odot}}
\def\fsun{\xi_{\odot}}
\def\zsun{{\rm Z}_{\odot}}
\def\msunyr{\msun {\rm yr}^{-1}}
\def\gdens{\msun\,{\rm kpc}^{-2}}
\def\sfrdens{\msun\,{\rm yr}^{-1}\,{\rm kpc}^{-2}}

\def\mum{\mu {\rm m}}
\newcommand{\angstrom}{\mbox{\normalfont\AA}}
\def\cc{\rm cm^{-3}}
\def\uflux{{\rm erg}\,{\rm s}^{-1} {\rm cm}^{-2} }

\def\fdust{\xi_{d}}
\def\fesc{f_{\rm esc}}
\def\td{\tau_{sd}}
\def\Sg{$\Sigma_{g}$}
\def\S*{$\Sigma_{\rm SFR}$}
\def\Ssfr{\Sigma_{\rm SFR}}
\def\Sgas{\Sigma_{\rm g}}
\def\Sstar{\Sigma_{\rm *}}
\def\Sesc{\Sigma_{\rm esc}}
\def\Srad{\Sigma_{\rm rad}}
\def\sSFR{\rm sSFR}
\def\tff{t_{\rm ff}}

\def\Dsolar{${\cal D}/\dsun$}
\def\Zsolar{$Z/\zsun$}
\def\DDsolar{\left( {{\cal D}\over \dsun} \right)}
\def\ZZsolar{\left( {Z \over \zsun} \right)}
\def\kms{{\rm km\,s}^{-1}\,}
\def\skms{$\sigma_{\rm kms}\,$}
\def\Mpc2{\ M_\odot\ \rm pc^{-2}}
\def\Mach{\mathbb{M}}

\def\lya{Ly$\alpha$}
\def\Scii{$\Sigma_{\rm [CII]}$}
\def\Sciimax{$\Sigma_{\rm [CII]}^{\rm max}$}
\def\CII{\hbox{[C~$\scriptstyle\rm II $]~}}
\def\CIII{\hbox{C~$\scriptstyle\rm III $]~}}
\def\OII{\hbox{[O~$\scriptstyle\rm II $]~}}
\def\OIII{\hbox{[O~$\scriptstyle\rm III $]~}}
\def\HH{\hbox{H$_2$}~} 
\def\HI{\hbox{H~$\scriptstyle\rm I\ $}} 
\def\HII{\hbox{H$\scriptstyle\rm II\ $}} 
\def\CIion{\hbox{C~$\scriptstyle\rm I $~}}
\def\CIIion{\hbox{C~$\scriptstyle\rm II $~}}
\def\CIIIion{\hbox{C~$\scriptstyle\rm III $~}}
\def\CIVion{\hbox{C~$\scriptstyle\rm IV $~}}
\def\nhh{n_{\rm H2}}
\def\nhi{n_{\rm HI}}
\def\nhii{n_{\rm HII}}
\def\fhh{x_{\rm H2}}
\def\fhi{x_{\rm HI}}
\def\fhii{x_{\rm HII}}
\def\fd{f^*_{\rm diss}} 
\def\ks{\kappa_{\rm s}}

\def\cyan{\color{cyan}}
\definecolor{apcolor}{HTML}{b3003b}
\definecolor{afcolor}{HTML}{800080}
\definecolor{lvcolor}{HTML}{DF7401}
\definecolor{mdcolor}{HTML}{01abdf} 
\definecolor{cbcolor}{HTML}{ff0000}
\definecolor{sccolor}{HTML}{cc5500} 
\definecolor{sgcolor}{HTML}{00cc7a}


\title{Is feedback-free star formation possible?}

\author[0000-0002-9400-7312]{A. Ferrara}
\email{andrea.ferrara@sns.it}
\author[0009-0009-1656-7769]{D. Manzoni}
\author[0000-0002-4324-0034]{E. Ntormousi}
\affil{Scuola Normale Superiore,  Piazza dei Cavalieri 7, 50126 Pisa, Italy}

\begin{abstract}
It has been suggested that, if the free-fall time of star-forming clouds is shorter than the lifetime ($\approx 3 $ Myr) of massive stars exploding as supernovae (SN), a large fraction of the cloud gas can be converted into stars during an allegedly `feedback-free' phase. Here, we show  that radiation pressure from Ly$\alpha$ photons produced in the pre-SN phase can instead erase feedback-free conditions, and severely limit the star formation efficiency (SFE). We find that, for a constant star formation rate, all clouds with gas surface density $(37-1.7 \times 10^5) \Mpc2$ have $\epsilon_* < 0.08$. Higher SFE values can only be reached if Ly$\alpha$-driven shells fragment and form stars. While advanced RHD simulations are required to establish the importance of this effect, adopting an optimistic guess, we find that the SFE increases with cloud surface density, rising from $\epsilon_*=0.023$ at $\Sigma_g = 37\ M_\odot\ \mathrm{pc^{-2}}$ to $\epsilon_*=0.27$ at $\Sigma_g = 1.7 \times 10^5\ M_\odot\ \mathrm{pc^{-2}}$. Given the optimistic assumptions adopted, these numbers should be regarded as upper limits. We conclude that Ly$\alpha$ radiation pressure strongly limits the SFE, even at solar metallicities, erasing the possibility that a feedback-free star formation mode with $\epsilon_* \simgt 0.4$ exists in the pre-SN phase. This conclusion remains valid even when other effects such as dust destruction of \lya\ photons, presence of \HII regions, velocity gradients, atomic recoil, and turbulence are considered.
\end{abstract}
\keywords{galaxies: high-redshift, galaxies: evolution, galaxies: formation, ISM: clouds, evolution}

\section{Introduction} \label{sec:intro}
The hydrogen Lyman-$\alpha$ (\lya) line is often the most prominent emission feature \citep{Osterbrock06} in the spectra of galaxies and has historically been used as a key tracer of star-forming systems \citep{Partridge67,Djorgovski92,Rhoads00,Taniguchi05,Ouchi09,Hu10,Pentericci11,Kashikawa11,Ouchi18,Shibuya18}. Surveys such as HETDEX \citep{Hetdex16}, VLT/MUSE \citep{MUSE19}, and SILVERRUSH \citep{Ouchi18,Shibuya18,Shibuya2019} have also revealed a significant population of Ly$\alpha$ emitters across a wide redshift range. 


In comparison, much less attention has been paid to the \emph{dynamical} role of \lya\ photons. As \lya\ often carries a significant fraction of the bolometric luminosity in young, metal-poor galaxies, it has long been suspected that its radiation pressure could significantly impact gas dynamics {\citep{Cox85, Bithell90}}. 

This physical effect might play a pivotal role in our understanding of the earliest galaxies now routinely investigated by the \textit{James Webb Space Telescope} (JWST). One of the major JWST findings so far has been the discovery of a population of surprisingly bright, blue galaxies at $z \gtrsim 10$ \citep{Naidu+2022, Harikane+23, McLeod+2024, Robertson+2024}. These sources, often termed “blue monsters,” exhibit ultraviolet luminosities and colors that challenge existing models of early galaxy formation \citep{Mason+2023, Mirocha&Furlanetto2023}. 

To reconcile observations with theory, several mechanisms have been proposed: extremely low dust attenuation \citep{Ferrara+2023}, top-heavy (or even PopIII-dominated \citealt{Maiolino+2024}) initial mass functions \citep{Trinca+2024, Schaerer+2024}, or modified cosmological initial conditions \citep{Liu&Bromm2022, Padmanabhan&Loeb2023}.

A particularly straightforward explanation suggests that such galaxies formed stars with extremely high efficiency, potentially converting nearly all their gas mass into stars during short-lived, `feedback-free' bursts  \citep[FFB,][]{Dekel+2023, Li+2023}. In this scenario, if the free-fall time of a star-forming molecular cloud is shorter than the delay before the first supernovae (typically $\sim 3$--$5$ Myr), star formation proceeds essentially unregulated, yielding star formation efficiencies (SFEs) approaching unity. Analytical arguments and radiation-hydrodynamic (RHD) simulations have claimed that early feedback processes—such as photoionization heating and UV radiation pressure on dust—are insufficient to halt star formation under these conditions \citep{Li+2023, Menon+2023}.

However, the FFB scenario neglects the momentum imparted by resonantly trapped \lya\ photons, which can become dynamically dominant well before the first SNe. At the extremely high neutral hydrogen column densities expected in compact, metal-poor systems, \lya\ photons undergo $\gg 10^6$ scatterings, with the resulting radiation pressure capable of disrupting gas clouds and quenching further star formation \citep{Tomaselli21, Kimm18, Nebrin+2024}. Crucially, this occurs on timescales shorter than the SN delay time, contradicting the key assumption of the FFB model.

As \lya\ often carries a significant fraction of the bolometric luminosity in young, metal-poor galaxies, it has long been suspected that its radiation pressure could impact gas dynamics \citep{Adams72,Harrington73,Neufeld90,Tan03,Oh:2001ex}. Early analytic and numerical studies suggested that trapped \lya\ photons could slow accretion onto protostars and dark matter halos, while Monte Carlo radiative transfer simulations showed that the momentum transfer from multiple scatterings can drive supersonic outflows \citep{Dijkstra08,Dijkstra09}. 

The dynamical influence ot \lya\ photons is often characterized by the \emph{force multiplier} $M_F \equiv \dot{p}_\alpha / (L_\alpha/c)$, which quantifies how photon trapping boosts the radiation force. In static, dust-free media, this multiplier scales as $M_F \propto \tau_0^{1/3}$, where $\tau_0$ is the \lya\ line-center optical depth \citep{Adams72,Smith17,Kimm18}. For typical star-forming clouds at high redshift, the optical depths are enormous ($\log \tau_0 \gtrsim 10$), leading to $M_F \sim 100$ or higher \citep{Tomaselli21, Nebrin+2024}, even in the presence of a moderate dust content \citep{Tomaselli21}.

The first hydrodynamical simulations explicitly incorporating \lya\ momentum coupling found that it can significantly alter the dynamics of early star-forming clouds. \citet{Smith:2006,Smith17,Smith18,Smith19} showed that \lya\ pressure can drive gas shell expansion faster than ionizing radiation alone, while \citet{Kimm18} implemented a subgrid model calibrated via MCRT simulations using the \textsc{RASCAS} code \citep{Michel20}, accounting for recoil, dust, and deuterium effects. Their results suggested that \lya\ feedback can potentially reduce star formation efficiency (SFE) and regulate cloud evolution prior to supernova (SN) explosions. Notably, $M_F$ saturates at high $\tau_0$ due to dust absorption, with values ranging from $\sim 50$ for solar metallicity to $\gtrsim 100$ in primordial environments.

Recent works have begun to highlight the regulatory role of \lya\ feedback. Using RHD simulations with subgrid \lya\ momentum coupling calibrated via Monte Carlo radiative transfer, \citet{Kimm+2018} showed that \lya\ pressure can reduce SFE and star cluster formation by factors of a few. Analytic models further indicate that \lya\ pressure alone can exceed the combined force from photoionization and UV continuum pressure by an order of magnitude \citep{Abe&Yajima2018, Tomaselli21, Nebrin25}. Notably, this regulation occurs well before the onset of SN explosions, fundamentally challenging the notion of a feedback-free star formation phase.

In this work, we argue that \lya\ radiation pressure is the dominant early feedback mechanism that regulates SFE in star-forming clouds. We study the evolution and overlap of bubbles whose expansion is driven by \lya\ radiation pressure. This feedback is capable of halting star formation and reducing the cloud's ability to convert gas into stars. Our model combines analytic estimates of \lya\ trapping and momentum coupling with a detailed treatment of cloud evolution under \lya-driven feedback.

The paper is organized as follows. In Sec. \ref{sec:bubbles} and Sec. \ref{sec:overlap} we compute the Ly$\alpha$-driven shell evolution and overlap. In Sec. \ref{sec:SFE} we use the model to determine the SFE of star-forming clouds when \lya\ feedback is included. Sec. \ref{sec:discuss} discusses additional effects due to dust destruction of \lya\ photons, the presence of \HII regions and other \lya\ radiative transfer aspects.  Sec. \ref{sec:summary} summarizes the paper. 

%
%
%
%
\begin{figure*}
\centering\includegraphics[width = 1.0 \textwidth]{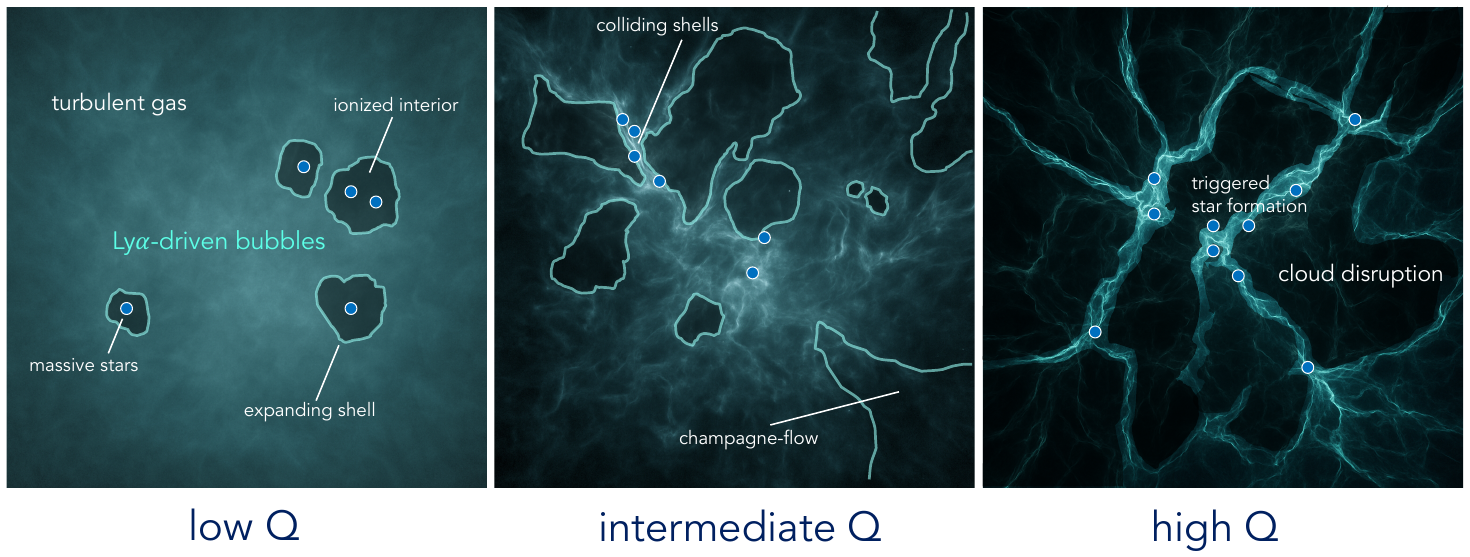}
\caption{  
Illustrative sketch of the Ly$\alpha$-driven bubble evolution in a star forming GMC. Ionized bubbles surrounded by neutral shells form around massive stars and expand due to \lya\ radiation pressure. While in the bubble interior star formation is suppressed, shells could become gravitationally unstable, fragment and form stars. As the volume filling factor $Q(t)$ increases, star formation is also facilitated by shell collisions in which the gas is highly compressed. Such triggered star formation accelerates the growth of bubbles until the remaining gas is either completely ionized or evacuated from the potential well of the GMC.  
}
\label{fig:sketch}
\end{figure*}

\section{Lyman Alpha-driven bubbles} \label{sec:bubbles}
Take a star-forming giant molecular cloud (GMC) of a given gas mass, $M_c$, and number density, $n = \rho/\mu m_p$, where $\mu=1.22$ is the mean molecular weight for an atomic neutral H+He mixture, and $m_p$ is the proton mass. Assuming the GMC to be in virial equilibrium, we define the virial parameter as
\begin{equation}\label{eq:avir}
    \alpha_{\rm vir} = \frac{5}{3} \frac{\sigma^2 R_c}{f G M_c} 
\end{equation}
where $f$ is a geometrical factor related to the cloud internal density profile. For spherical clouds with a radial density profile $\rho \propto r^{-\gamma}$, it is $f = (1-\gamma /3)/(1-2\gamma /5)$. We assume $f = 1$ for our homogeneous cloud and $\alpha_{\rm vir} = 5/3$, which is consistent with local observations
\citep{Heyer09} and simulations of GMCs \citep{Grisdale18}. From eq. \ref{eq:avir} we derive the expression for the cloud radius $R_c = \sigma t_{\rm ff}$, free-fall time $t_{\rm ff} = (3/4\pi G \rho)^{-1/2}$, and 1D velocity dispersion $\sigma = (GM_c/t_{\rm ff})^{1/3}$. Finally, for some purposes the following expression for the gas surface density, $\Sigma_g = M_c/\pi R_c^2$, is useful:
\begin{equation}\label{eq:sigmag}
    \Sigma_g = 800 \left(\frac{M_c}{10^6\ M_\odot}\right)^{1/3} \left(\frac{n}{10^3\ \rm cm^{-3}}\right)^{2/3}\ M_\odot\ \rm pc^{-2}
\end{equation}

If one writes the mean star formation rate in the cloud using the standard Schmidt-type law, then using the previous expressions we find
\begin{equation}\label{eq:SFR}
    {\rm SFR}_0 = \epsilon_{\rm ff} \frac{M_c}{t_{\rm ff}} = \epsilon_{\rm ff}\frac{\sigma^3}{G}.
\end{equation}
In the following we take the 'instantaneous' efficiency per free-fall time as $\epsilon_{\rm ff}=0.01$ based on the average value of local measurements obtained by \cite{Krumholz07}.

As soon as they form, massive stars (mass $m_*$), produce H-ionizing photons at a rate $\dot {N}_\gamma$. Part of these photons are converted by recombinations into Ly$\alpha$ photons exerting a radiation pressure onto the surrounding gas.  As a result, the gas is compressed in an expanding thin shell enclosing a low-density bubble, whose radius, $R_s$, increases with time according to the solution of the momentum equation,
\begin{equation}\label{eq:momentum}
    \frac{d}{dt}[M(<R_s)\dot R_s] = M_F(R_s) \frac{L_\alpha}{c}
\end{equation}
Interestingly, this mechanism works even in metal-free gas as it does not rely on the presence of dust grains (or free electrons, in the standard Eddington case) to transfer the radiation momentum to the gas. In eq. \ref{eq:momentum} we have neglected the gravity force, $F_g = G m_* M_s/R_s$ exerted by the star on the shell, which can be shown to be negligible before shells start to overlap. If individual bubbles are small with respect to the cloud radius $R_s \ll R_c$, the tidal effects of the global cloud gravity field can also be neglected.

The Ly$\alpha$ luminosity\footnote{A negligible LyC escape fraction  from the GMC is assumed.}, $L_\alpha = (2/3) E_a \dot {N}_\gamma$, where $E_\alpha = 10.2$ eV is the energy gap of the transition, drives the radiation pressure. The ionization rate depends on stellar mass and metallicity. We adopt a metallicity\footnote{We note that $\dot {N}_\gamma$ is only weakly dependent on metallicity, varying by a factor $\simlt 2$ in $0.01 < Z/Z_\odot < 1$\label{fn:dust}.} consistent with that measured in early galaxies ($\approx 1/50\ Z_\odot$) and use the results in \citet{Schaerer02}:
\begin{equation}
    \log \dot {N}_\gamma = 27.80 + 30.68 x - 14.80 x^2 + 2.50 x^3,
\end{equation}
with $x = \log(m_*/M_\odot)$. For a $1-100\ M_\odot$ Salpeter IMF which we adopt here, the IMF-weighted mean ionizing luminosity of massive ($m_* > 8 M_\odot$) stars is $\dot {N}_\gamma = 10^{48.7}\ \rm s^{-1}$ corresponding to a characteristic stellar mass $\hat m_* = 26.6\ M_\odot$. The previous value of $\dot {N}_\gamma$ implies $L_\alpha = 1.3\times10^4 L_\odot = 5\times 10^{37}\ \rm erg\ s^{-1}$. Note that $L_\alpha$ would be even larger for a top-heavy IMF.

The force is amplified by the resonant nature of the Ly$\alpha$ scattering, resulting in the aforementioned force multiplier, $M_F$. Physically, $M_F$ represents the ratio of the trapping time (due to multiple scatterings) of Ly$\alpha$ photons to the light crossing time in a system of characteristic size $\ell$: $M_F=t_{\rm trap}/({\ell}/c)$. In the absence of dust (see below), the force multiplier can be written, following \cite{Tomaselli21} {(see also \citealt{Lao&Smith2020})}, as
\begin{equation}\label{eq:MF}
    M_F\approx3.51\,(a_v \tau_0)^{1/3},
\end{equation}
where $\tau_0 =\sigma_0 N_{\rm HI}$ is the optical depth at the line center, $\sigma_0=5.88\times 10^{-14} T_4^{-1/2} \rm cm^{2}$, $T=30$ K is the adopted gas temperature\footnote{We use the notation $Y_X = Y/10^X$.}, and $N_{\rm HI}$ is the neutral hydrogen column density collected by the shell, on which radiation pressure acts; finally, $a_v=4.7\times10^{-4}T_4^{-1/2}$. Here, we set $N_{\rm HI} = n R_s$. 
{We note that eq. \ref{eq:MF} is valid for a central point source, which should be appropriate in the context of this study. }

If dust is present, radiation pressure is limited by the fact that \lya\ photons are absorbed by grains, thus decreasing the force multiplier $M_F$. To account for this effect we use the derivation by \citet{Tomaselli21}, and cap $M_F$ above a certain $\tau_0$ where dust absorption becomes important. This is equivalent to imposing the following condition:
\be\label{eq:MF_minimum}
M_F = \min[M_F, M_F(D)]
\ee
where $M_F$ is given by eq. \ref{eq:MF}, and
\be\label{eq:MF_limiter}
M_F(D) = 35.2\ (T_4 D)^{-1/4},
\ee
and $D$ is the dust-to-gas ratio normalized to the Milky Way value (1/162), which we take to be proportional to the metallicity, $D \propto Z$. 
{Expression eq. \ref{eq:MF_minimum} is also in excellent agreement with the $M_F$ predictions obtained by \citet{Nebrin+2024} (see their Fig. 5) from a novel analytical \lya\ radiative transfer solution that includes the effects of continuum absorption, gas velocity gradients, \lya\ destruction, ISM turbulence and atomic recoil. }

Let us introduce the non-dimensional variables $y = R_s/R_c$, $\dot y = \dot R_s/\sigma$, and $\tau = t/t_{\rm ff}$. We also write $M_F = 3.51\,(a_v \sigma_0 n R_c)^{1/3} x^{1/3} \equiv M_0 x^{1/3}$. Eq. \ref{eq:momentum} becomes:
\begin{equation}\label{eq:momentum_norm}
    \frac{d}{d\tau}[y^3 \dot y] = {\cal K}_\alpha y^{1/3},
\end{equation}
where ${\cal K}_\alpha(n, \sigma) = M_0 G L_\alpha/\sigma^4 c$ is the non-dimensional radiation pressure force coefficient. The solution of eq. \ref{eq:momentum_norm} shows that the shell radius increases as a power-law function of time:
\begin{equation}\label{eq:rs}
    y(\tau) = \left(\frac{121 {\cal K}_\alpha}{78}\right)^{3/11} \tau^{6/11}.
\end{equation}

The expansion described by eq. \ref{eq:rs} continues until the stars explode as SNe at the end of their life. Using the fits provided in \citet{Raiteri96}, the lifetime of a $\hat m_* = 26.6\ M_\odot$ star at $Z=1/50\ Z_\odot$ is $t_* = 6.6$ Myr. As the main goal of this paper is to study the pre-SN feedback phase and assess whether radiation pressure from massive stars can disperse the cloud and limit star formation, we concentrate in the following on evolutionary times $t \le t_*$. 

%
%
%
%
\begin{figure*}
\centering\includegraphics[width = 1.0 \textwidth]{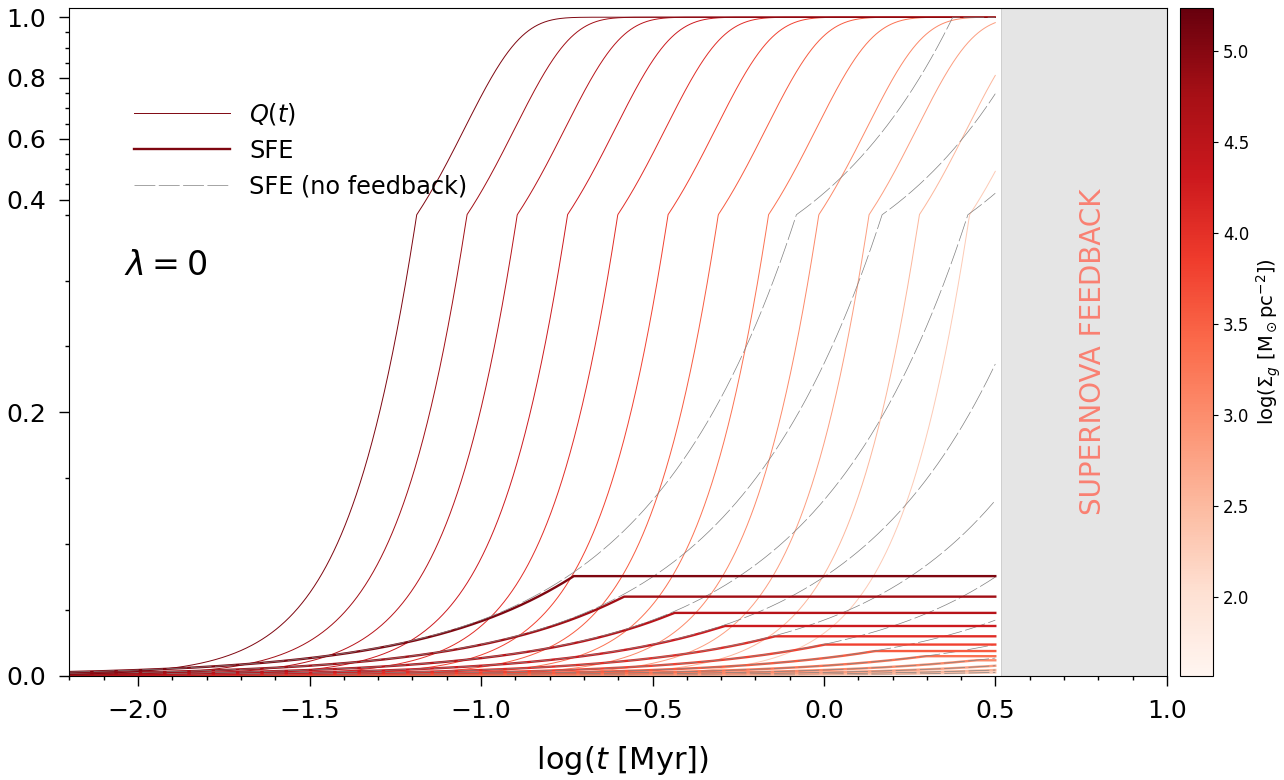}
\caption{  
Time evolution of Ly$\alpha$-driven bubbles volume filling factor $Q$ (thin red lines) and star formation efficiency (SFE), $\epsilon_* = M_*/M_c$ (thick red curves) up to the onset of SN explosions at $t=3$ Myr for a constant star formation rate ($\lambda=0$, see Sec. \ref{sec:res_const}). The curves are color-coded according to the cloud gas surface density, $\Sigma_g$, as shown in the colorbar. Also shown (gray dashed lines) is the SFE expected for each cloud in the absence of Ly$\alpha$ feedback. Note that the vertical axis has been expanded for display purposes.
}
\label{fig:Q_const}
\end{figure*}

\section{Bubble overlap} \label{sec:overlap}
Armed with the solution for the growth of bubbles around individual stars, we now concentrate on their collective behaviour. The volume filling factor, $Q(t)$, i.e. the fraction of the GMC volume filled with bubbles, is determined by the following differential equation: 
\begin{equation}\label{eq:Q_ode}
    \frac{dQ}{dt} =  \nu\ {\rm SFR}\ \frac{V_s(t)}{V_c}(1-Q).
\end{equation}
The $(1-Q)$ term on the r.h.s. accounts for bubble overlapping, $V_s/V_c = (R_s/R_c)^3 = y^3$, and $\nu = 1/52.89$ is the number of massive ($>8\ M_\odot$) stars per unit stellar mass formed appropriate for the adopted IMF. We first solve eq. \ref{eq:Q_ode} imposing a constant star formation rate ${\rm SFR}_0$ (eq. \ref{eq:SFR}); we then generalize the result to cases in which the SFR is itself a function of $Q$.  

If we neglect overlapping, and we denote with $Q_0(t)$ the solution in this case, we find
\begin{equation}\label{eq:Q_no_ovl}
    Q_0(t) = \nu\frac{\rm SFR_0}{V_c}  \int_0^{t} V_s(t-t') dt', 
\end{equation}
which in terms of normalized variables becomes
\begin{equation}\label{eq:Q_no_ovl_norm}
    Q_0(\tau) = N_* \left[\frac{11}{29} \left( \frac{121 {\cal K}_\alpha}{78}\right)^{9/11} \tau^{29/11} \right], 
\end{equation}
where $N_* = (\nu\ {\rm SFR_0}\ t_{\rm ff}$) is the number of massive stars formed per free-fall time.

The previous formula is valid as long as the filling factor is low ($Q \simlt 0.3$). For larger values, bubble overlapping becomes important, so the solution of eq. \ref{eq:Q_ode} takes the form
\begin{equation}\label{eq:Q}
    Q(\tau) = 1 - e^{-Q_0(\tau)}.
\end{equation}
As the Ly$\alpha$-driven bubbles expand, they carve low-density, ionized bubbles surrounded by dense shells. The probability that, at time $\tau$, a given point in the GMC is located within a bubble, is given by $Q(\tau)$. 

%
%
%
%
\begin{figure*}
\centering\includegraphics[width = 1.0 \textwidth]{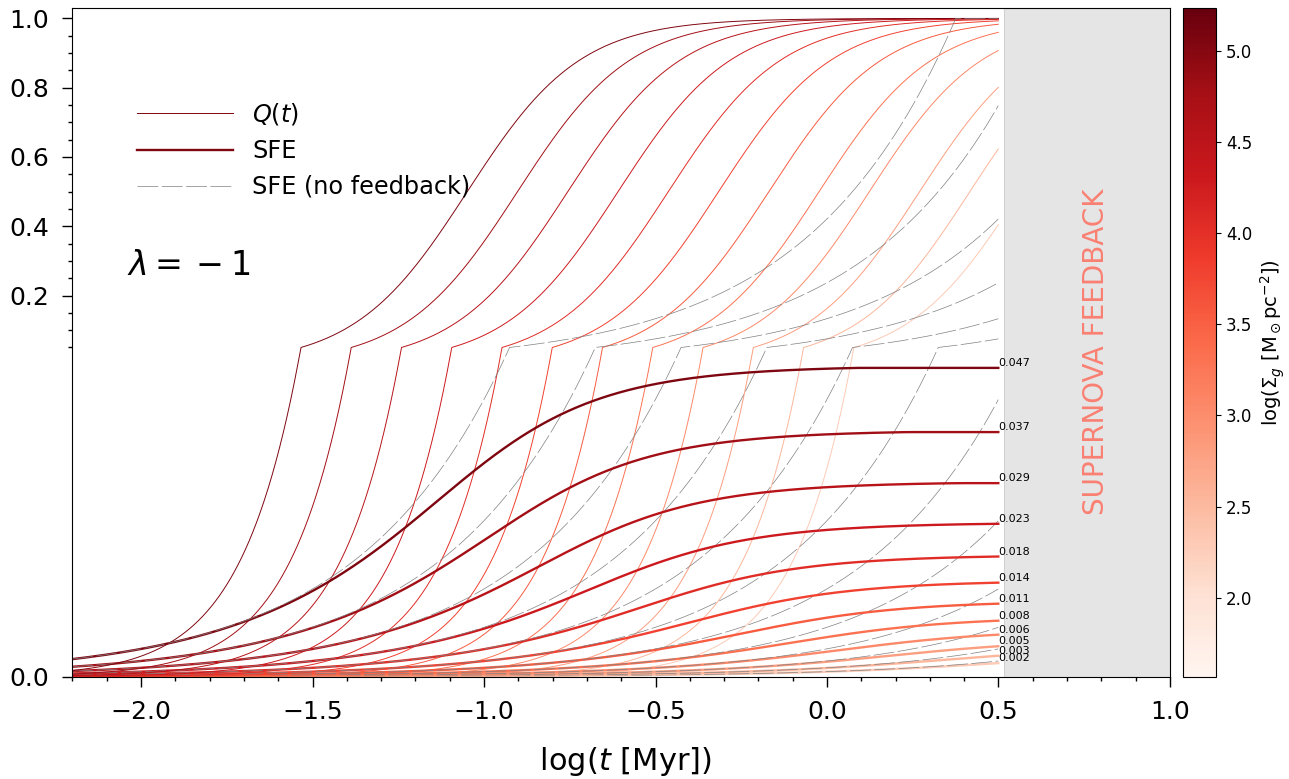}
\caption{  
As Fig. \ref{fig:Q_const} but including the negative feedback of Ly$\alpha$-driven bubbles on the SFR ($\lambda = -1$, see Sec. \ref{sec:res_negfeed}). The numbers indicate the final value of the SFE, $\epsilon_* = M_*/M_c$, just before the onset of SN explosions at $t=3$ Myr. 
}
\label{fig:Q_var_lambda-1}
\end{figure*}

The above treatment assumes that the SFR is constant. This is equivalent to neglecting the feedback of bubbles on the star formation process. This feedback can be both negative or positive. Negative feedback occurs because the bubble interior contains low-density, ionized gas where star formation is virtually impossible. Hence, the overall SFR should be decreased by a factor $(1-Q)$. On the other hand, gas collected in the expanding shells can become gravitationally unstable, fragment and form stars (see sketch in Fig. \ref{fig:sketch}). This process represents a positive feedback on the SFR, whose strength increases with $Q$. 

To describe this behaviour we write the SFR in the following generic form:
\begin{equation}\label{eq:variable_SFR}
    {\rm SFR} = {\rm SFR}_0 (1+\lambda Q), 
\end{equation}
where $\lambda$ is an arbitrary constant. Note that $\lambda \ge -1$ in order to avoid an unphysical negative SFR for large $Q$. If $\lambda < 0$, negative feedback dominates, and as $Q$ increases, star formation is  suppressed. If instead $\lambda > 0$, the positive feedback produces a SFR increase. Finally, the special case in which negative and positive feedback exactly balance each other (or the presence of bubbles is ignored) corresponds to the case $\lambda=0$ with a solution given by eq. \ref{eq:Q}. Physically-motivated values of $\lambda$ will be discussed in the following Section. 

To obtain the general solution for any value of $\lambda \ge -1$, we substitute eq. \ref{eq:variable_SFR} into eq. \ref{eq:Q_ode} and solve for $Q(t)$:
\begin{align}\label{eq:Q_varSFR_combined}
Q(\tau) = 
\begin{cases}
    \dfrac{e^{(1+\lambda) Q_0(\tau)} - 1}{\lambda + e^{(1+\lambda) Q_0(\tau)}}, & \text{if } \lambda > -1, \\[1em]
    \dfrac{Q_0(\tau)}{1 + Q_0(\tau)}, & \text{if } \lambda = -1.
\end{cases}
\end{align}
Note that if $\lambda = 0$ we recover the constant SFR case described by eq. \ref{eq:Q}. In the following we will use this solution to discuss the fate of the GMC under the action of Ly$\alpha$ radiation pressure feedback.

\section{Star formation efficiency} \label{sec:SFE}
Using the formalism developed in the previous Sections, we want to determine the value of the gas-to-stars conversion factor, $\epsilon_* = M_*/M_c$, where $M_*$ is the total mass of stars formed before SN explosions take place at $t=t_*$. We will derive this quantity as a function of the cloud properties. We explore two cases corresponding to either constant (eq. \ref{eq:SFR}) or evolving (eq. \ref{eq:variable_SFR}) SFR. 

\subsection{Constant SFR}\label{sec:res_const}
The constant SFR case is obtained by setting $\lambda = 0$ in eqs. \ref{eq:variable_SFR}--\ref{eq:Q_varSFR_combined}, and therefore $\rm SFR = SFR_0$. The key features for this case are illustrated in Fig. \ref{fig:Q_const} for clouds with $\Sigma_g$ in the range\footnote{In practice, we fix $M_c=10^6\ M_\odot$ and vary $n$ to obtain the desired range.} $(37-1.7\times 10^5)\ M_\odot\ \rm pc^{-2}$ bracketing the observed and predicted range for GMCs in different environments. 

The filling factor $Q$ grows with time for all clouds. As $Q$ increases, a larger fraction of the GMC volume is filled with low-density, ionized gas in which SF is not possible. Eventually, SF stops when $Q \approx 1$. We define the time at which SF is quenched as $t_Q$. We note that this treatment is not completely self-consistent as the SFR should in principle decrease with time as the filling factor increases. 

The condition $Q=1$ is only reached by clouds with $\Sigma_g \gtrsim 100 \Mpc2$ before SNe start to explode at $t_{\rm SN} = 3$ Myr (roughly the lifetime of the most massive stars). Due to their higher SFR, the most massive clouds reach $Q=1$ already at $t_Q \approx 0.2$ Myr. 

Next, we compute the stellar mass formed, $M_*$, by time-integrating the SFR for each cloud; this mass is removed from the available gas mass. We then define the cloud star formation efficiency as $\epsilon_* = M_*/M_c$. The evolution of $\epsilon_*(t)$ is shown by the thick red curves in Fig. \ref{fig:Q_const}. As we see, feedback from Ly$\alpha$ radiation pressure limits the SFE to very low values ($\epsilon_* < 0.08$), ending SF well well before 3 Myr. 

Although the constant SFR is a special (and unlikely) case in which positive and negative feedback exactly balance, it allows a direct comparison with the standard feedback-free star formation model, if we further remove the condition that SF is completely quenched once $Q=1$. For comparison, we also show (gray dashed lines) the SFE expected for each cloud in the absence of Ly$\alpha$ feedback. Indeed, clouds with $\Sigma_g > 4\times 10^4 \Mpc2$ can reach a SFE $\epsilon_* > 0.4$ before SNe occur. The most massive cloud, with its $\rm SFR = 0.43\  \msunyr$, is able to transform all its gas into stars in merely 2.4 Myr. This is in striking contrast with the results obtained when Ly$\alpha$ feedback is considered. Thus, if Ly$\alpha$ feedback is included, the potential feedback-free star formation mode yielding large SFE values is essentially erased. The SF suppression can be even more drastic during the pre-SN phase, as we are neglecting here the effects of \HII regions. We'll return to this point in Sec. \ref{sec:HII}.

\subsection{Evolving SFR (negative feedback only)}\label{sec:res_negfeed}
As we have noticed in Sec. \ref{sec:overlap}, Ly$\alpha$ feedback can either quench or boost star formation depending on $\lambda$. We first discuss the negative feedback-only case, in which SFR is suppressed within bubbles. This case corresponds to $\lambda=-1$ and it is shown in Fig. \ref{fig:Q_var_lambda-1}.

The emerging picture is not fundamentally different from the constant SFR case analyzed above. Now the SFR decreases with time and this slows
down the growth of $Q(t)$. While in principle the system has more time available to form stars before the cloud is dispersed at $Q=1$, this advantage is counterbalanced by the slower conversion of gas into stars. As a result the final value of $\epsilon_*$ is roughly the same as in the constant SFR case.

From Fig. \ref{fig:Q_var_lambda-1} we see that the condition $Q=1$ is now reached by the most massive clouds with  $\Sigma_g \gtrsim 8000 \Mpc2$ before SNe start to explode. In these clouds the SFE is $\epsilon_* < 0.048$ (see the individual values shown in the Figure; also note that the $y$-axis scale has been expanded for a better display). Less massive clouds continue to form stars up to $3$ Myr as their SFR is too low to produce a strong Ly$\alpha$ feedback; however, and for the same reason, their ability to convert gas into stars is very modest ($\simlt 1$\%). 

We conclude that -- unless Ly$\alpha$-driven bubbles produce some sort of positive feedback in addition to the negative one -- there is no pre-SN feedback-free phase in which the SFE exceeds $5-10$\%.

%
%
%
%
\begin{figure*}
\centering\includegraphics[width = 1.0 \textwidth]{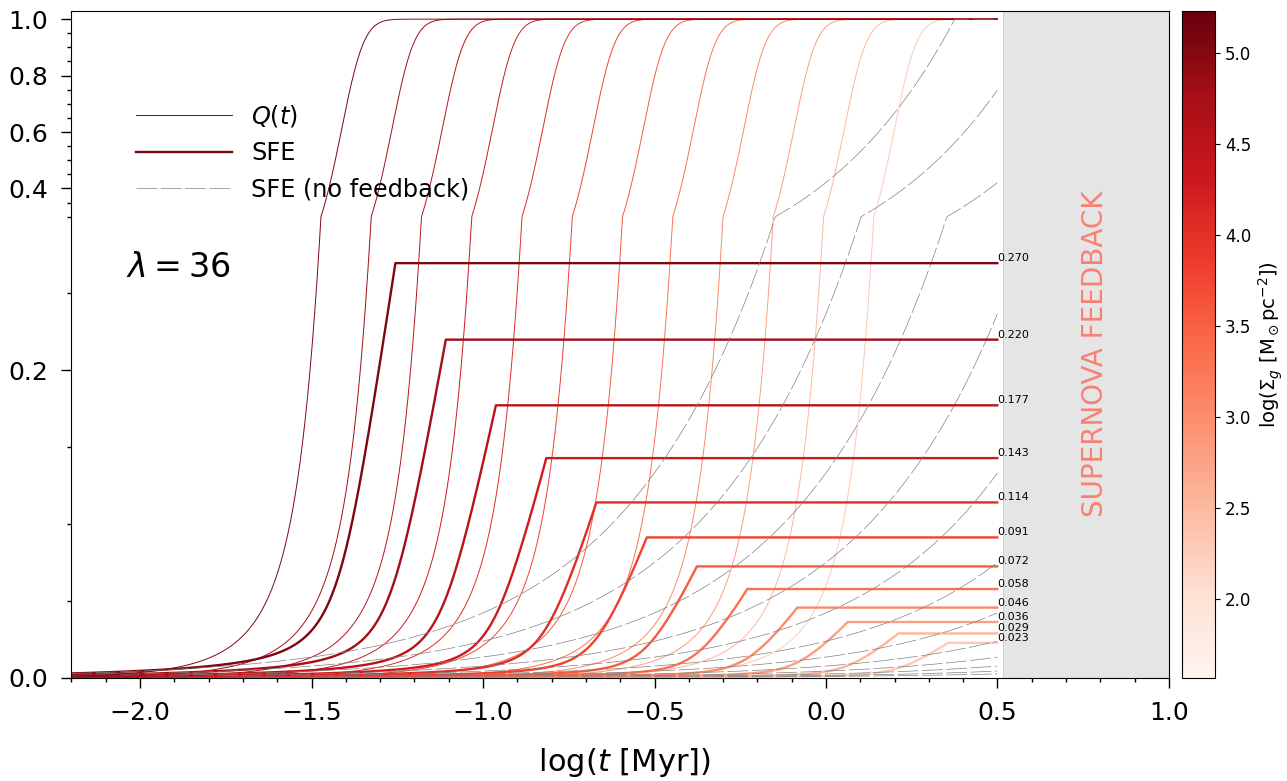}
\caption{  
As Fig. \ref{fig:Q_const} but including both negative and positive feedback of Ly$\alpha$-driven bubbles on the SFR (we show results for the `optimistic' case $\lambda = 36$, see Sec. \ref{sec:res_posfeed}). The numbers indicate the final value of the SFE, $\epsilon_* = M_*/M_c$, just before the onset of SN explosions at $t=3$ Myr. 
}
\label{fig:Q_var_lambda36}
\end{figure*}

\subsection{Evolving SFR with triggered star formation}\label{sec:res_posfeed}
The treatment in Sec. \ref{sec:res_const} assumes that SFR is completely suppressed inside bubbles (negative feedback). While this remains a motivated assumption, it neglects the possibility that SFR can be instead enhanced (positive feedback) in swept-up shells and in the web of filaments produced by their collisions, a picture preliminary discussed in Sec. \ref{sec:overlap} (see sketch in Fig. \ref{fig:sketch}). If positive feedback overcomes the negative one, the feedback parameter $\lambda$ becomes positive too. 

While radiation-hydrodynamical simulations including the dynamical treatment of Ly$\alpha$ radiation are necessary to fully characterize the importance of such triggered (or self-propagating) star formation process, some preliminary, educated guess of the $\lambda$ value can nevertheless be made. 

If the shells become gravitationally unstable, fragment and form stars, from eq. \ref{eq:SFR} we see that the SFR enhancement over the mean value $\rm SFR_0$ is controlled by three processes: (a) the fraction of the shell mass that ends up into gravitationally-bound fragments, $\delta M_s$, (b) the density enhancement, $\delta n$, of the shell gas with respect to the mean, (c) the star formation efficiency per free-fall time enhancement, $\delta \epsilon_{\rm ff}$, with respect to the global value $\epsilon_{\rm ff}$. Thus, we can estimate $\lambda$ from the relation 
\begin{equation}
\lambda \simeq \delta M_s \times \delta\epsilon_{\rm ff} \times (\delta n)^{1/2},
\label{eq:lambda}
\end{equation}
where the square-root dependence on $\delta n$ comes from the free-fall time in eq. \ref{eq:SFR}.  Let us qualitatively evaluate these three factors.

Simulations of SN--driven and \HII region--driven shells \citep{Dale09, Walch13} typically find that, for thin shells, the fraction of mass ending up in bound clumps is \(\delta M_s \approx 0.1\). For an isothermal strong shock, the post-shock density enhancement scales with the Mach number as \(\delta n = \mathcal{M}^2\). In the case of Ly\(\alpha\)-driven shells, the velocity evolves according to the derivative of eq.~\ref{eq:rs}, \(\dot{y} \propto \tau^{-5/11}\). At \(t \approx 3\) Myr, depending on the initial gas surface density \(\Sigma_g\), we find \(\mathcal{M} \approx 1{-}15\). To maximize the star-formation enhancement, we adopt the upper value \(\mathcal{M} = 15\), corresponding to \((\delta n)^{1/2} = 15\).  

Clumps produced by shell fragmentation are expected to be dense and compact \citep{decataldo:2019}, allowing them to self-shield efficiently against UV radiation and resist gas ablation. Their star formation efficiency could therefore exceed that of the parent cloud \citep{Lada10, Krumholz19}.  \citet{Murray11}, by analyzing a sample of 32 star-forming GMCs in the Milky Way, found that while for the cloud-averaged value of $\epsilon_{\rm ff}$ is in the range is 0.002-0.2 (consistent with our choice $\epsilon_{\rm ff} = 0.01$), for individual clumps, $\epsilon_{\rm ff}$ raises to $0.14-0.24$. Adopting again the maximum value, we find $\delta\epsilon_{\rm ff}=24$. Inserting these values into eq.~\ref{eq:lambda} yields an optimistic estimate of \(\lambda = 36\). 
 
The results for $\lambda = 36$ are shown in Fig.~\ref{fig:Q_var_lambda36}. This choice corresponds to a scenario in which the SFR is overwhelmingly dominated (36:1) by triggered star formation resulting from Ly$\alpha$-driven shell fragmentation. The final star formation efficiency prior to the onset of SN feedback increases with cloud surface density, rising from $\epsilon_*=0.023$ at $\Sigma_g = 37\ M_\odot\ \mathrm{pc^{-2}}$ to $\epsilon_*=0.27$ at $\Sigma_g = 1.7 \times 10^5\ M_\odot\ \mathrm{pc^{-2}}$. Given the optimistic assumptions adopted, these numbers should be regarded as upper limits. For comparison, the more conservative value $\lambda = 14$ yields $0.015 < \epsilon_* < 0.186$ over the same $\Sigma_g$ range.  

The relatively low final SFE is primarily driven by the rapid suppression of star formation caused by the swift expansion of Ly$\alpha$-driven bubbles, which ionize and/or expel gas from the GMC, thereby removing it from the star formation cycle. Under such sustained SFR, the bubbles fill the entire GMC volume before the onset of SN explosions; in the most massive clouds, this occurs in  $\simlt 10^5\ \mathrm{yr}$.  

We therefore conclude that Ly$\alpha$ radiation pressure imposes a stringent limit on the fraction of GMC gas that can be converted into stars during the feedback-free pre-SN phase.

%
%
%
%
\begin{figure*}
\centering\includegraphics[width = 1.0 \textwidth]{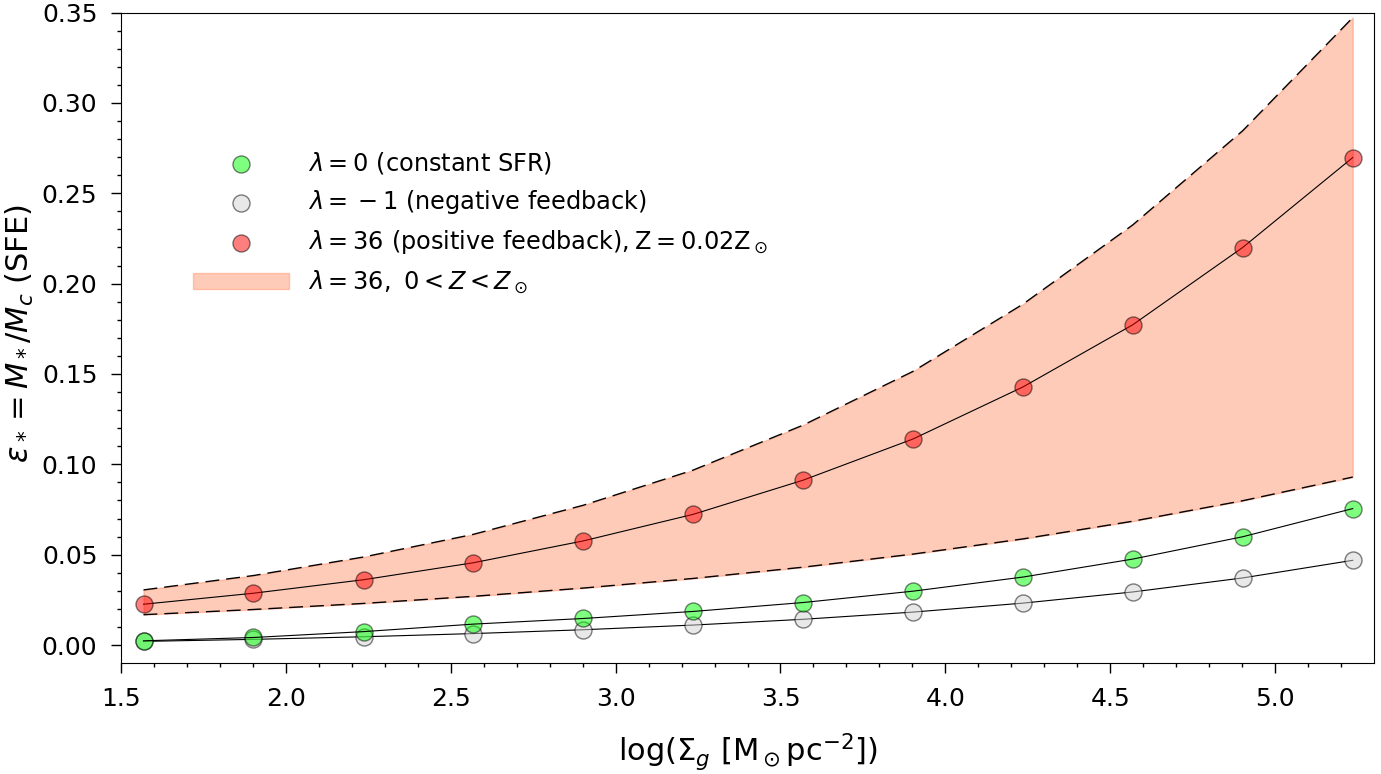}
\caption{  
Summary plot showing the predicted dependence of the SFE of star-forming clouds on the cloud gas surface density, $\Sigma_g$, including the effects of \lya\ feedback. We show the three cases corresponding to different values of the feedback parameter $\lambda = 0$ (green circles, constant SFR), $\lambda = -1$ (grey circles, including negative feedback only), $\lambda = 36$ (red circles, including negative and optimistic positive feedback). For the latter case we show the impact of varying the gas metallicity in the range $0<Z< Z_\odot$ (coral shaded area) as discussed in Sec. \ref{sec:dust}.      
}
\label{fig:SFE_summary}
\end{figure*}

\section{Discussion} \label{sec:discuss}
The simple yet robust model we have developed shows that the SFE of molecular clouds in the pre-SN
phase is regulated by an often ignored physical process, Ly$\alpha$ radiation pressure from 
young, massive stars. As low density, ionized bubbles grow and fill an increasingly large fraction, $Q$ 
of the cloud volume, star formation is suppressed in such cavities, but could be triggered in the boundary shells
which fragment into dense and opaque clumps. 

Describing this process in detail is difficult, and dedicated, high-resolution RHD simulations in which \lya\ dynamics is properly included are necessary. Although there a few early and promising attempts in this direction none of these studies can provide a definite answer yet due to the lack of one of the key ingredients.
While pioneering studies \citep{Dijkstra&Loeb2009}, corroborated by the most recent and advanced analytical models
\citep{Tomaselli21, Nebrin+2024, Smith+2025}, have indisputably shown the importance of Ly$\alpha$ radiation pressure
feedback in various environments, and most notably at high redshifts, numerical simulations are still 
falling short of providing quantitative and detailed predictions. 

In fact, state-of-the-art multi-physics RHD galaxy simulations, such as e.g. SERRA \citep{Pallottini22} or FIRE \citep{Hopkins20}, 
include various radiative feedback channels (photoionization, dust radiation pressure, etc.) but historically 
did not include, mostly due to their computational cost, on-the-fly \lya\ radiation transfer and associated dynamical effects. 

This gap is beginning to be filled by some studies \citep{Smith18, Kimm18, Michel20}. In particular, \citet{Smith18} explored 
on-the-fly \lya\ RHD in 1D/idealized setups and developed algorithms (rDDMC / resonant DDMC) to massively 
accelerate \lya\ RT so that it can be coupled to hydrodynamics in 3D in the near future. Once fully implemented, these simulations
will represent perfect follow-up experiments to test the present results, pin-point uncertain processes, and assess the role of 
triggered star formation.

The results found here are summarized in Fig. \ref{fig:SFE_summary}. There we show the relation between the final SFE $\epsilon_*$
at the end of the pre-SN phase (taken here to be 3 Myr) as a function of the cloud surface density $\Sigma_g$. We do confirm an increasing
trend of the SFE $\epsilon_*$ in more massive clouds. This behaviour has been already suggested by previous authors \citep{Li23, Menon25, Somerville25} for \HII regions- and SN-driven feedback. The SFE follows with good accuracy a power-law of the type $\epsilon_*(\Sigma_g, \lambda) = A(\lambda) \Sigma_g^\gamma$, with $\gamma \simeq 0.3$. Interestingly, the feedback does not modify the power-law index $\gamma$; rather, it controls the normalization factor $A$ via the feedback parameter $\lambda$. For the three cases explored here we find $A(\lambda) = (1.2,1.9,7.2)\times 10^{-3}$ for $\lambda = (-1,0,36)$, respectively.

\subsection{Dust effects}\label{sec:dust}
The results discussed so far have been obtained for a gas metallicity $Z = 0.02 Z_\odot$. In this context, metallicity controls two quantities: (a) the stellar ionizing photon production rate, $\dot N_{\gamma}$, and (b) the dust-to-gas ratio, $D$, which we assume to scale linearly with $Z$. Concerning (a) we have already noted (see footnote \ref{fn:dust}) that the the effect is negligible. The dust abundance is instead more critical and we discuss it in the following.

Dust, by absorbing \lya\ photons, can decrease the value of $M_F$ below the dust-free case (eqs. \ref{eq:MF} and \ref{eq:MF_limiter}). Although the reduction depends weakly on $D$, it does affect the results. To illustrate this point we have rerun the positive feedback $\lambda=36$, for a range of metallicities, spanning the range $0 < Z < Z_\odot$. The outcome is shown by the coral shaded area in Fig. \ref{fig:SFE_summary}. For solar metallicity (top boundary of the area) a reduced $M_F$ value allows a 27\% increase of the SFE from the fiducial case with $Z=0.02 Z_\odot$. For a metal free gas though, the SFE is strongly suppressed, and it is about 3 times lower ($\epsilon_* = 0.09$). 

Thus, \lya\ radiation pressure effectively limits star formation even for solar metallicities. {Surprisingly, though, many important works have neglected \lya\ radiation pressure at solar metallicity, e.g. \citet{Arthur96, Draine11}.} Moreover, we have to note that our study does not include direct radiation pressure on dust grains, which is mainly due to the much more numerous non-ionizing photons. 
Clearly, as $Z$ increases, this pressure force becomes more important. Including this extra term is beyond the purpose of the present work; for reference, the relative importance of \lya\ and dust-mediated radiation pressure on the gas has been discussed in detail in \citet{Tomaselli21}.  However, it is clear that dust, if present, might be another factor in erasing the alleged feedback-free star formation pre-SN phase by providing and extra radiation pressure channel.    
{We stress again that advanced RHD simulations including dust dynamics are required to test our results under the idealized geometry adopted. For example, \citet{Nebrin25} showed that dust destruction of \lya\ photons can be stronger for extended sources.} 

\subsection{HII regions}\label{sec:HII}
So far, we have neglected the effects of \HII regions around massive stars. Qualitatively, we expect that over-pressurized, ionized regions also drive expanding bubbles in the GMC, similar to the Ly$\alpha$-driven bubbles studied here. Therefore, neglecting \HII regions likely leads to an \textit{overestimate} of the SFE. As we have seen, $\epsilon_* \propto \Sigma_g^{0.3}$, meaning the most massive and dense clouds are those attaining a relatively high SFE (e.g., above $10\%$). From Fig. \ref{fig:SFE_summary}, taking the most favourable case $\lambda=36$, this occurs for $\Sigma_g \simgt 10^4 \Mpc2$.

To illustrate the point, let us examine the most massive cloud in our range, with $\Sigma_g = 1.7 \times 10^5\ \Mpc2$ and a corresponding density of $n = 3 \times 10^6\ \cc$ (eq. \ref{eq:sigmag}). For the fiducial value $\log \dot N_\gamma = 48.6$, the Strömgren radius, which delimits the ionized region, is $R_I = 2.5 \times 10^{-3}$ pc. The ionization front (IF) reaches this radius in approximately a recombination time, $t_r = (\alpha_B n)^{-1} \approx 0.04$ yr. By that time, the size of the \lya\ bubble (from eq. \ref{eq:rs}) would be $R_s = (1/16) R_I$. Thus, Ly$\alpha$ driving acts on the neutral layer just beyond $R_s$. In other words, the \HII region provides a `kick-start' to the Ly$\alpha$-driven bubble.  

Once the transition to a D-type IF occurs, the bubble begins to expand dynamically, driven by the internal pressure of the ionized gas. This expansion generates a shock in the surrounding neutral gas. Assuming an isothermal shock, and uniform, photoionization-equilibrium conditions in the ionized region, the velocity of the IF at radius $R$ can be expressed as \citep{Sommovigo20}:
\begin{equation}\label{eq:IF}
    \frac{1}{c_{\rm I}} \frac{dR}{dt} = \left(\frac{R_{\rm I}}{R}\right)^{3/4} - C\left(\frac{R}{R_{\rm I}}\right)^{3/4},
\end{equation}
where $C = (c_{\rm s}^{2} + \sigma^{2})/(c_{\rm I}^{2} + \sigma^{2})$, and $c_{I}$ and $c_{\rm s}$ are the sound speeds in the ionized and neutral gas, respectively. With the boundary condition $r(t=0) = R_{\rm I}$, eq. \ref{eq:IF} can be integrated analytically. However, for our purposes, it suffices to determine the maximum (stalling) radius, $R_{I, \rm max}$, of the expanding \HII region. Setting $dR/dt = 0$ yields:
\begin{equation}\label{eq:RImax}
    R_{I, \rm max} \simeq R_I \left[ 1 + \left(\frac{c_I^2}{\sigma^2}\right)\right]^{2/3}.
\end{equation}
The expansion stalls at $R_{I, \rm max}$ once the internal pressure is balanced by the external turbulent pressure, $\rho \sigma^2$. In eq. \ref{eq:RImax}, we have also used the fact that $c_s \ll c_I$.  

For our most massive molecular cloud, taking $c_I = 10\ \kms$, we find $c_I/\sigma = (10/56.2) \simeq 0.18$. From eq. \ref{eq:RImax}, this implies that $R_{I, \rm max}$ is only 2\% larger than $R_I$ -- meaning the \HII region barely expands beyond $R_I$.  

We conclude that, in the absence of \lya\ radiation pressure, \HII regions would have a negligible impact on star formation and its efficiency of the most massive clouds, given their extremely low volume filling factor. The influence of \HII regions becomes more significant for smaller clouds. However, this is largely irrelevant here, as such clouds already exhibit low $\epsilon_*$ values.  

These preliminary conclusions must be verified through detailed numerical simulations that include the evolution of IFs and the combined effects of \lya\ and dust-mediated radiation pressure. Nevertheless, our results provide robust upper limits on the SFE of star-forming clouds in the pre-SN phase.

\subsection{Other neglected effects}\label{sec:other}
We have not made any attempt to include the effects of velocity gradients, atomic recoil, or turbulent density fluctuations on the force multiplier, $M_F$. While the first two processes have been shown \citep{Nebrin+2024} to be largely subdominant\footnote{{Using their analytical model solutions, \citet{Nebrin25} have shown that for the point source geometry of interest here atomic recoil and velocity gradients do not appreciably modify the $M_F$ trend in eq. \ref{eq:MF}. An additional effect, destruction of \lya\ photons by $2p \to 2s$ transitions, might become important at high $\Sigma_g$ thus mimicking dust effects in pristine environments.}} with respect to \lya\ photon destruction due to dust (see Sec. \ref{sec:dust}), some authors \citep[e.g.][]{Munirov23} have claimed that turbulent fluctuations with a finite correlation length, $\ell_t$, could in principle reduce the average number of scatterings (and therefore $M_F$) suffered by \lya\ photons before escaping the cloud. This effect becomes important when the correlation length is small, i.e. when the ratio $\ell_t/\lambda_{\rm mfp} \simlt 10^4$, where $\lambda_{\rm mfp} = (n\sigma_0)^{-1}$ is the mean free path of a \lya\ photon at the line centre. The above relation can be translated in a condition on $\ell_t < 0.05/n\ \rm pc$. As $\ell_t$ is measured to be $10-100$ pc in GMCs, for any density value considered here, turbulence should not have any significant effects on $M_F$. This conclusion is also supported by the results in \citet{Nebrin+2024}. 

Finally, we recall that we have not included the energy input due to stellar winds from massive stars. Although the cumulative energy per unit stellar mass formed is only $\approx 1\%$ of that in ionizing radiation (see Fig. 2 of \citealt{Pallottini17}), the more efficient winds energy coupling with the gas can partly compensate for the mismatch. Stellar winds could represent yet another, although maybe subdominant, SFE limiting factor during the first 3 Myr.

\section{Summary} \label{sec:summary}
We have investigated whether \lya\ radiation pressure from young, massive stars is effective in 
limiting the conversion of gas into stars (i.e. the star formation efficiency, SFE) in their parent  
molecular clouds prior to the onset of SN explosions ($\approx 3$ Myr from the beginning of star formation). 
The main goal of the study is to assess whether an early feedback-free evolutionary phase during 
which star formation occurs almost unimpeded by feedback processes and reach high SFE values.

To this aim we have developed a simple model describing the evolution and overlap of Ly$\alpha$-driven bubbles, 
until star formation is quenched at a time $t_Q$ corresponding to a bubble volume filling factor, $Q \simeq 1$, 
when the cloud gas is either fully ionized and/or evacuated. Our study examines a wide range of 
cloud gas surface densities, $\Sigma_g = 37-1.7\times 10^5\ M_\odot\ \rm pc^{-2}$ bracketing the observed and 
predicted range for GMCs in different environments. The main results are: 

\begin{itemize}
\item[{\color{red} $\blacksquare$}] If \lya\ bubbles do not feed back on the star formation rate ($\rm SFR \approx \rm const.$ or feedback parameter $\lambda=0$ in eq. \ref{eq:Q_varSFR_combined}), Ly$\alpha$ radiation pressure limits the SFE to very low values ($\epsilon_* < 0.08$) independently of $\Sigma_g$, quenching the formation of new stars well before 3 Myr. This result is in stark contrast with the standard `feedback free' model predictions, according to which clouds with $\Sigma_g > 4\times 10^4 \Mpc2$ reach a SFE $\epsilon_* = 0.4-1.0$ before SNe occur.

\item[{\color{red} $\blacksquare$}] The impact of \lya\ radiation pressure on the SFE is even more dramatic if we account for the decrease of the SFR within Ly$\alpha$-driven bubbles ($\lambda=-1$, negative feedback). The condition $Q=1$ is reached at $t < 3$ Myr by clouds with $\Sigma_g \gtrsim 8000 \Mpc2$; in these systems, $\epsilon_* < 0.048$. Less massive clouds continue to form stars up to 3 Myr as their SFR is too low to produce a strong Ly$\alpha$ feedback; however, and for the same reason, their SFE is very modest ($\simlt 1\%$). 

\item[{\color{red} $\blacksquare$}] Higher SFE values can only be attained if Ly$\alpha$-driven shells fragment and form stars (triggered star formation) thus causing an increase in the SFR ($\lambda>0$, positive feedback). While advanced RHD simulations are necessary to determine the value of $\lambda$, based on the physical arguments given, we argue that $\lambda \simlt 36$. Adopting this value as an optimistic guess, we find that the SFE increases with cloud surface density, rising from $\epsilon_*=0.023$ at $\Sigma_g = 37\ M_\odot\ \mathrm{pc^{-2}}$ to $\epsilon_*=0.27$ at $\Sigma_g = 1.7 \times 10^5\ M_\odot\ \mathrm{pc^{-2}}$. Given the optimistic assumptions adopted, these numbers should be regarded as upper limits. 

\item[{\color{red} $\blacksquare$}] We conclude that Ly$\alpha$ radiation pressure strongly limits the fraction of GMC gas that can be converted into stars, essentially erasing the possibility that a genuine feedback-free star formation mode with $\epsilon_* \simgt 0.4$ exists in the pre-SN phase. 

\item[{\color{red} $\blacksquare$}] Our conclusion remains valid even when (i) the dust/metal content of the cloud is varied from metal-free to solar values, (ii) we allow for the presence of \HII regions adding another negative feedback on the SFR, (iii) the effects of velocity gradients, atomic recoil, and turbulent density fluctuations on the \lya\ force multiplier are considered. 
\end{itemize}

\acknowledgments
We thank A. Smith for useful discussions. 
This work is supported by the ERC Advanced Grant INTERSTELLAR H2020/740120, and in part by grant NSF PHY-2309135 to the Kavli Institute for Theoretical Physics. 
This work made use of the MCMC sampler EMCEE \citep[][]{EMCEE13}.
Plots are produced with the \textsc{matplotlib} \citep{Hunter07} package.

\bibliographystyle{aasjournal}
\bibliography{paper}
\end{document}